# Robust Contextual Outlier Detection: Where Context Meets Sparsity

Jiongqian Liang and Srinivasan Parthasarathy
Computer Science and Engineering, The Ohio State University, Columbus, OH, USA
{liangji,srini}@cse.ohio-state.edu

## ABSTRACT

Outlier detection is a fundamental data science task with applications ranging from data cleaning to network security. Given the fundamental nature of the task, this has been the subject of much research. Recently, a new class of outlier detection algorithms has emerged, called *contextual outlier detection*, and has shown improved performance when studying anomalous behavior in a specific context. However, as we point out in this article, such approaches have limited applicability in situations where the context is sparse (i.e., lacking a suitable frame of reference). Moreover, approaches developed to date do not scale to large datasets. To address these problems, here we propose a novel and robust approach alternative to the state-of-the-art called RObust Contextual Outlier Detection (`ROCOD`). We utilize a local and global behavioral model based on the relevant contexts, which is then integrated in a natural and robust fashion. We also present several optimizations to improve the scalability of the approach. We run `ROCOD` on both synthetic and real-world datasets and demonstrate that it outperforms other competitive baselines on the axes of efficacy and efficiency (40X speedup compared to modern contextual outlier detection methods). We also drill down and perform a fine-grained analysis to shed light on the rationale for the performance gains of `ROCOD` and reveal its effectiveness when handling objects with sparse contexts.

## 1. INTRODUCTION

"An *outlier* is an observation that deviates so much from other observations as to arouse suspicion that it was generated by a different mechanism" [16]. Detecting outliers finds applications in a wide range of domains including cyber-intrusion detection, epidemiology studies, fraud detection, data cleaning and textual anomaly detection. As described in a recent survey [9], there has been a plethora of work seeking to detect outliers from the statistics, geometry, machine learning, database, and data mining communities.

A number of efforts in this space have treated all attributes, associated with a data point, in an egalitarian fashion. However, in many domains, some attributes are usually highly related to the outlier behavior, called *behavioral attributes* or *indicator attributes*, while other attributes only provide contexts of the behavior, termed *contextual attributes*. It has been demonstrated recently that by distinguishing contextual attributes from behavioral attributes, the precision of outlier detection can be increased [9, 33, 34, 37, 14, 15]. Formally, *contextual outlier* or *conditional anomaly* is defined as an object with behavior deviating from other objects with similar contextual information [13, 9, 33, 34]. Usually, contextual attributes are used to define the contexts, and objects sharing similar contexts with an object form its *reference group*. Behavior attributes, on the other hand, are used for examining outlierness in a specific context, compared to the reference group.

One pitfall of existing contextual outlier detection methods is that they might fail to examine the outlierness of objects with sparse contexts. To intuitively show this, we use a toy example of credit card fraud detection. For simplicity, suppose we intend to detect suspicious transactions and only monitor two variables, the annual income of cardholders and the amount of each transaction. Figure 1 is a scatter plot of all the data points. $x$-axis and $y$-axis represent the two variables that are monitored.

**The Importance of Contextual Attributes**: Though contextual attributes are not directly related to the anomalous behavior, they provide useful information on contexts for outlier detection. In the example of Figure 1, transaction amount and annual income can be respectively regarded as the behavioral attribute and contextual attribute. If we merely consider the behavioral attribute, then points G, E and F will not be flagged as outliers, which is not reasonable. Therefore, we need auxiliary information from the contextual attributes to pinpoint the outliers.

**Incorporating Contextual Attributes**: One conventional way to incorporate contextual attributes is treating them similarly with behavioral attributes by concatenating the two together. In the example above, all the boundary points (A, B, ..., G) will be reported as outliers. Another way is to use existing contextual outlier detection techniques, such as [33]. Following the definition of the contextual outlier, we can examine the difference of behavioral attribute between a point and other points with the similar contextual attribute. By using this approach, one will report point C

.

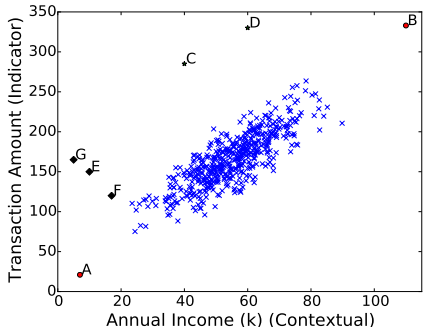

Figure 1: Toy example of contextual outliers.

and D as outliers since their behavior values ($y$ values) are quite different from other points that contain similar contextual values ($x$ values).

**Limitations of Existing Approaches**: These two methods have limitations on addressing objects with sparse contexts (A, B, E, F and G). The former approach tends to overestimate the outlierness of objects with unusual contexts because the outlierness score can be directly contributed by contextual attributes. Moreover, the latter approach fails to properly examine the outlierness of objects with sparse contexts due to the lack of reliable reference groups. In fact, applying state-of-the-art contextual outlier detection algorithm [33] for the toy example, we obtain outlierness score ranking of $B > D > C > A \gg G, E, F$. However, a close look at our example reveals that A and B should ideally not be flagged as outliers since they follow the normal pattern between the two attributes that a lower value in contextual attribute corresponds to a lower value in behavioral attribute and vice versa. Intuitively, customers with higher annual income are more likely to spend more money in one transaction. Therefore, we need a more robust approach for outlier detection to distinguish E, F, G with A and B, giving A and B lower outlierness scores. To the best of our knowledge, this paper is the first piece of work attempting to rectify this problem.

To this end, we propose a refined approach to better exploit contextual attributes to assist outlier detection, called RObust Contextual Outlier Detection (`ROCOD`). `ROCOD` can particularly address the problems caused by the contextual sparsity, making it more robust towards broad outlier detection tasks. Specifically, we propose *local expected behavior* and *global expected behavior* models that seek to understand the dependence structure between behavioral attributes and contextual attributes. Local expected behavior models are designed to predict the behavior by referring to the objects with similar context, called *contextual neighbors*. Global expected behavior models learn the dependence structure between contextual attributes and behavioral attributes from the data, and infer the behavior in a holistic sense. We then propose a regularized integration function to naturally couple both types of behavior models based on the number of contextual neighbors, which naturally accommodates objects with sparse contexts. We subsequently develop some novel optimizations (by leveraging locality sensitive hashing (LSH) [10, 18, 11] to detect contextual neighbors) and parallelization strategies to scale up our algorithm to large datasets. Finally, we run `ROCOD` on both synthetic and real-world datasets and compare it with *five* state-of-the-art outlier detection techniques. Our experimental results show that `ROCOD` outperforms all the baselines on both effectiveness (measured by different metrics) and efficiency. A detailed drill-down analysis reveals that the better performance of `ROCOD` is attributed to its robustness on addressing objects with sparse contexts.

To summarize, our contributions are:

- Identify and tackle a novel problem in existing contextual outlier detection techniques that cannot deal with objects containing abnormal contextual attributes.
- Model the expected behavior of each object from both local and global perspectives, which are subsequently regularized in a natural way.
- Significantly scale up our algorithm by leveraging LSH and parallel computing paradigms.

## 2. RELATED WORK

Outlier detection or anomaly detection has been widely studied for decades. There are many off-the-shelf surveys, review papers, and books on this topic [9, 26]. Essentially, existing outlier detection techniques can be roughly divided into global approaches and local approaches.

Global approaches usually assume the data follows a certain kind of statistical distribution and measure the outlierness score of objects using metrics related to the model [3, 31]. A representative method in statistics is to model the data using Gaussian Mixture Models (GMM) [31]. The outlierness score of each object is usually measured by the Mahalanobis distance to the mean of the mixture model or simply the probability density of the object under the distribution. Another category of outlier detection techniques, local approaches, usually determine the outlierness score by comparing the objects to the local reference group. Distance-based outlier detection approaches study the distance to the neighborhoods and use it to measure the abnormality of an object [7, 30, 20, 21]. $k$-NN method is a typical distance-based method, which uses the distance to the $k$-th nearest neighbor as the outlierness score. Density-based methods compare the density of an object with the neighbors and objects with obviously lower density are more likely to be outliers [6, 35]. However, all these approaches usually combine contextual and behavioral attributes, simply assuming they contribute equally to indicating outlier behavior.

More related work is subspace outlier detection [2, 42, 22]. They dealt with high dimensional data and tried to project the original data into lower dimensional space [2] or focused on finding the outlying subspaces. [42]. Kriegel *et al.* [22] studied the local correlations of attributes for outlier detection and found outliers in arbitrary subspaces based on Principal Component Analysis (PCA). However, these approaches cannot be directly applied to contextual outlier detection scenario since the subspace or principal component they output might be mainly contributed by the contextual attributes. Without distinguishing contextual attributes from behavior ones, these methods are likely to generate undesired results.

Contextual outlier detection have been studied particularly in time-series data [40, 32], spatial data [25], and spatio-temporal data [23]. In these specific problems, context is temporal, spatial or spatio-temporal attributes. For example, spatial outliers are defined as objects whose non-spatial attributes are different from their spatial neighbors. As a special case of contextual outlier detection, they deal with simple and fixed contextual attributes (spatial attributes) and their methods usually cannot be generalized to other applications where the context can be broader areas.

On the other hand, general contextual outlier detection is relatively new and has not been studied until recent years. Song et al. [33] proposed a statistical approach to detect outliers assuming that behavioral attributes conditionally depend on the environmental or contextual attributes. They employed GMM to model both contextual attributes and behavioral attributes, and used a mapping function to capture their probabilistic dependence. EM is then adopted to estimate the parameters of the model. Similarly, Hong et al. [17] model the data distribution by a multi-dimensional function, based on which conditional outlierness scores are assigned to each object. One drawback of these approaches is that they are not scalable to large dataset since it is computationally expensive to learn the model.

Valko et al. [37] proposed a non-parametric graph-based algorithm to perform conditional anomaly detection. Starting with a labeled training set, the algorithm conducts label propagation in the graph and estimates the confidence of labeling. Some domain-specific approaches were also proposed for different applications [14, 15]. But all these methods assumed labeled data is available, which is not true for most of the real world applications. Wang et al. [38] addressed the problem of detecting contextual outliers in graphs using random walk, which is not applicable to general dataset without graph structures. Tang et al. [34] operated on categorical relational data and used data cube computation techniques to discover contextual outliers.

To the best of our knowledge, none of the above works addresses the problem caused by the sparsity of contexts. In this paper, we intend to tackle this problem by utilizing both local and global approaches, which are fused in a way depending on the size of the reference group.

## 3. ROBUST CONTEXTUAL OUTLIER DETECTION (ROCOD)

In this section, we introduce our RObust Contextual Outlier Detection (ROCOD) approach in detail.

### 3.1 Problem Formulation

Given a series of objects, the $i$-th object can be represented as
$$z^{(i)} = \begin{pmatrix} x^{(i)} \\ y^{(i)} \end{pmatrix} = (x_1^{(i)}, x_2^{(i)}, ..., x_C^{(i)}, y_1^{(i)}, y_2^{(i)}, ..., y_B^{(i)})^T \quad (1)$$

where $z^{(i)}$ is the whole attribute vector, $x^{(i)}$ is contextual attribute vector and $y^{(i)}$ is behavioral attribute vector. Without loss of generality, we assume $x^{(i)}$ has $C$ dimensions and $y^{(i)}$ has $B$ dimensions. Following this, the whole dataset can be denoted as $Z = \langle z^{(1)}, z^{(2)}, ..., z^{(N)} \rangle$, where $N$ is the total number of objects. Among them, we denote $O$ as the set of outliers. Given the dataset $Z$, our goal is to assign each object $i$ with an outlierness score $S_i$ so that outliers in $O$ have much higher values than other objects.

Conforming to the definition of the contextual outlier, the outlierness of an object arises from the abnormal behavioral attributes in its particular context. In other words, provided the contextual attributes, there is underlying pattern restricting the behavioral attributes to some expected values, beyond which one object will be considered as an outlier. We here call it the *dependent pattern* and define the expected behavior as follows.

DEFINITION 1. **Expected Behavior.** *For object $i$, the expected behavior is the values of its behavioral attributes predicted by the underpinning dependent pattern, given its contextual attributes $x^{(i)}$.*

Formally, denote the underpinning dependent pattern as a function $f(\cdot)$, then the expected behavior is
$$\hat{y}^{(i)} = f(x^{(i)}) \quad (2)$$

Following the definition, a contextual outlier is the object with behavioral attributes violating the dependent pattern under its contextual attributes. We can gauge the outlierness score of object $i$ by measuring the difference of $\hat{y}^{(i)}$ and $y^{(i)}$. Therefore, revealing the expected behavior of each object is a crucial part of flagging contextual outliers.

However, it is nontrivial to find out the pattern function $f(\cdot)$ and the expected behavior since the data distribution is usually unknown and the data is inherently noisy. We next discuss how we approach this problem from both a local and global perspective and how to combine them.

### 3.2 Local Expected Behavior Modeling

We first study the dependent pattern and expected behavior from the local aspect. We define *contextual neighbors*:

DEFINITION 2. **Contextual Neighbors.** *Contextual neighbors of an object are the objects that are similar to it on contextual attributes.*

Formally, the set of contextual neighbors of object $i$ is
$$CN_i = \left\{ j : j \in D \wedge j \neq i \wedge sim(x^{(i)}, x^{(j)}) \geq \alpha \right\} \quad (3)$$

where $\alpha$ is a predefined similarity threshold and $sim(\cdot)$ is a similarity function of two vectors. $D = \{1, 2, ..., N\}$ denotes the set of objects indexes. While $sim(\cdot)$ can be any reasonable similarity function, we choose cosine similarity here.

We then define *local expected behavior* as follows.

DEFINITION 3. **Local Expected Behavior.** *The local expected behavior of an object is the average values of behavioral attributes among all its contextual neighbors.*

This definition hinges on the underlying assumption of contextual outlier detection that objects with similar contextual attributes share similar behavioral attributes [9, 33, 34] and is also a natural extension of Tobler's first law of geography [36]. To formalize it, the expected behavior of object $i$ given contextual attributes $x^{(i)}$ is
$$\hat{y}^{(i)} = \Phi(x^{(i)}) = \frac{\sum_{j \in CN_i} y^{(j)}}{|CN_i|} \quad (4)$$

where $\Phi(\cdot)$ denotes local behavior pattern.

The local behavior pattern is tightly tied to the definition of contextual outlier detection and is supposed to directly reflect the dependent relationship between behavioral attributes and contextual attributes. Moreover, it does not make any assumption on the distribution of data. While this local property has been widely used in spatial and temporal outlier detection [40, 25], we generalize it to a broader range of applications with arbitrarily defined contexts.

However, since the local expected behavior relies on the contextual neighbors, it will be inapplicable if one object does not have contextual neighbors. If an object shares little contextual information with others, the number of its contextual neighbors will be very few or even zero. This is a challenging problem in contextual outlier detection. As we pointed out in Section 1, the object lacks a set of reference to define the expected behavior when the contextual attributes are sparse. Therefore, the local expected behavior cannot be inferred for all the objects and we need a more robust way to compute the expected behavior.

### 3.3 Global Expected Behavior Modeling

We now introduce a global approach to capture the underlying dependent pattern in the data and infer the expected behavior, called *global expected behavior*. Instead of merely using local contextual neighbors, we seek for a holistic approach that leverages all the objects to predict the expected behavior.

A natural way of capturing the global dependent relationship between behavioral attributes and contextual attributes is to adopt regression models. For each behavioral attribute, we can learn a regression model considering contextual attributes as features and the behavioral attributes as the target values. In total, we learn $B$ regressors from the dataset with $B$ behavioral attributes. With the regression models, we hereby define global expected behavior.

DEFINITION 4. **Global Expected Behavior.** *The global expected behavior of an object is the values of behavioral attributes predicted by the regression models taking its contextual attributes as input.*

Formally, the global expected behavior of object $i$ is

$$\hat{y}^{(i)} = \Psi(x^{(i)}) \quad (5)$$

where $\Psi(\cdot)$ incorporates the regression models learned from the whole dataset using contextual attributes as independent variables and behavioral attributes as dependent variables. $\Psi(\cdot)$ takes the contextual attributes $x^{(i)}$ as input and outputs the expected behavior attribute vector $\hat{y}^{(i)}$. Note that we can adopt any regression model here, either linear or non-linear. Obviously, this manner of behavior modeling is a holistic approach since the regression models are learned from the whole dataset.

### 3.4 Ensemble Expected Behavior

So far we have introduced two different perspectives to depict the underpinning relationship between behavioral attributes and contextual attributes, and used them to compute local and global expected behavior. Local expected behavior adopts the contextual neighbors as the reference frame while global expected behavior is predicted by adopting the regression models learned from the whole dataset.

These two ways of inferring expected behavior offer complementary benefits. Local expected behavior is a model-free approach and infers the expected value in a localized fashion. When the number of contextual neighbors is large, the local expected behavior should have a lower bias. However, local expected behavior contains larger variance in general and is prone to noise, especially when the number of contextual neighbors is small. In fact, when there is no contextual neighbor, it cannot be applied at all. On the other hand, global expected behavior infers the values considering the dependent relationship between the two categories of attributes in a holistic way. Therefore, it is expected to contain smaller variance and is more robust against noise. However, its bias tends to be larger since it cannot capture the fine-grained local pattern for each object.

Considering the different advantages of these two approaches, we intend to find an appropriate manner to integrate them. One possible method to combine these two approaches is to simply take a regularized (weighted) sum of them and use it to infer the expected behavior. One can set the weights of these two methods beforehand and apply them to all the objects. However, it is not trivial to determine reasonable weights since we do not have evidence on which metric is more important. Moreover, the issue of zero contextual neighbors still remains, making it inapplicable to infer local expected behavior for some objects.

To resolve these issues, we propose an adaptive weighted sum to integrate these two methods. Instead of fixing the weights for all objects, we adjust the weights according to the number of contextual neighbors of each object. Given an object $i$, we define the ensemble expected behavior as

$$\hat{y}^{(i)} = \lambda_i \cdot \Phi(x^{(i)}) + (1 - \lambda_i) \cdot \Psi(x^{(i)}) \quad (6)$$

where

$$\lambda_i = \frac{\sqrt{|CN_i|}}{\max_{1 \leq j \leq N} \sqrt{|CN_j|}}. \quad (7)$$

Here, $\Phi(x^{(i)})$ and $\Psi(x^{(i)})$ are respectively local expected behavior and global expected behavior as defined in Equation 4 and Equation 5. The intuition behind this weighted combination is that if an object has a sufficient number of contextual neighbors, we believe the contextual neighbors are a reliable reference frame and place more weight on local expected behavior. Otherwise, we put more weight on the global metric. In addition, we take the square root transformation on $|CN_i|$ to improve the normality of $\lambda_i \cdot \Phi(x^{(i)})$ [4].

By setting weight on local expected behavior proportional to the square root of the number of contextual neighbors, the model is more robust and can appropriately deal with context sparsity. In particular, it properly solves the problem caused by zero contextual neighbors by naturally setting the weight of local expected behavior to zero and putting all the weight on global expected behavior when $|CN_i| = 0$.

With the ensemble expected behavior for each object, we measure the outlierness score of object $i$ by the difference of expected behavior $\hat{y}^{(i)}$ and real behavior $y^{(i)}$, specifically by the L2-norm $\left\| y^{(i)} - \hat{y}^{(i)} \right\|_2$. Here, the L2-norm of difference assumes each behavioral attribute contributes equally to the outlierness score. However, this might not be true since the attributes can have different credibility at flagging the outlier. For example, if the real values of one behavioral attribute are highly consistent with the expected ones, indicating the pattern is well captured by our approach as a whole, then a slight difference of real value and expected value should be a strong sign of outlierness. Following this intuition, we weight each behavioral attribute based on how good the expected behavior on capturing the real behavior values. For each behavioral attribute, we use the coefficient of determination [27] to measure consistency between the real behavior and the expected one. For $j$-th behavioral

attribute, the coefficient of determination is calculated by

$$R^2(y_j, \hat{y}_j) = 1 - \frac{\sum_{1 \leq i \leq N}(y_j^{(i)} - \hat{y}_j^{(i)})^2}{\sum_{1 \leq i \leq N}(y_j^{(i)} - \bar{y}_j)^2} \quad (8)$$

where $\bar{y}_j = \frac{1}{N}\sum_{1 \leq i \leq N} y_j^{(i)}$, $y_j^{(i)}$ is the value of $j$-th behavioral attribute for object $i$, and $\hat{y}_j^{(i)}$ is the expected value of it using our approach. We define the weight of $j$-th behavioral attribute as $w_j = \max(R^2(y_j, \hat{y}_j), 0)$ and therefore the range of $w_j$ is $[0, 1]$. We now can calculate the outlierness score of object $i$ using

$$S_i = \left\| W^T(y^{(i)} - \hat{y}^{(i)}) \right\|_2 \quad (9)$$

where $W = (w_1, w_2, ..., w_B)^T$.

ROCOD detects outliers using Equation 9. Specifically, ROCOD uses it to capture the outlierness score for each object and the $n$ objects with highest outlierness scores are selected as outliers, where $n$ is the predetermined number of outliers the user is interested in.

## 3.5 Scaling Up the Algorithm

### 3.5.1 Finding Contextual Neighbors Efficiently

To compute the local expected behavior, we need to find out all the contextual neighbors of each object, i.e., all the objects that are similar to it with cosine similarity higher than $\alpha$ on contextual attributes. Here we employ a methodology based on Locality Sensitive Hashing (LSH) for cosine similarity (a natural measure given our methodology) to generate candidates [10, 8]. The basic idea of LSH is to apply hash functions so that similar objects can generate the same hash values with high probability. Specifically, corresponding to a random vector $r$, each dimension of which is drawn from a standard normal distribution, for two vectors $x^{(i)}$ and $x^{(j)}$, we have

$$\mathbf{Pr}[\text{sign}(r \cdot x^{(i)}) = \text{sign}(r \cdot x^{(j)})] = 1 - \frac{\theta(x^{(i)}, x^{(j)})}{\pi}. \quad (10)$$

where $\text{sign}(\cdot)$ is the sign function and $\theta(\cdot)$ is the angle between the two vectors. Here, $h_r(x) = \text{sign}(r \cdot x)$ is a locality-sensitive hash function corresponding to specific random vector $r$ geared towards cosine similarity. Then given a threshold $\alpha$, if two objects with attributes $x^{(i)}$ and $x^{(j)}$ satisfy $\cos(x^{(i)}, x^{(j)}) \geq \alpha$, we have

$$\mathbf{Pr}[\text{sign}(r \cdot x^{(i)}) = \text{sign}(r \cdot x^{(j)})] \geq 1 - \frac{\arccos \alpha}{\pi}. \quad (11)$$

Denote $1 - \frac{\arccos \alpha}{\pi}$ as $p$ and recall as $\epsilon$. In order to attain recall $\epsilon$, we need $l$ signatures for each object, each of which is a concatenation of $d$ hashes. We can derive that $l = \lceil \frac{\log(1-\epsilon)}{\log(1-p^d)} \rceil$. In this paper, we set $\epsilon$ to 0.975 and fix $d$ to 8. After applying the hash functions to the whole dataset, two objects become candidates of contextual neighbors for each other if they at least have one signature in common. We generate contextual neighbor candidates for each object in the dataset and then examine whether their exact cosine similarity is larger than the similarity threshold.

Moreover, we point out that cosine similarity can be estimated by simply examining the fraction of the generated hash values that two objects agree upon [8], which will further alleviate the computation. Specifically, for an object $i$ and its contextual neighbor candidate object $j$, concatenate their $l$ signatures respectively to form binary numbers with a length of $l * d$, denoted as $sig_i$ and $sig_j$. The concatenated signatures are essentially a binary number with length of $l * d$, where $d$ is the length of one single signature and is set as 8 in this paper. Applying Equation 10, we have

$$1 - \frac{\theta(x^{(i)}, x^{(j)})}{\pi} \approx \frac{l * d - Ham(sig_i \oplus sig_j)}{l * d} \quad (12)$$

Then, instead of exactly examining $\cos(x^{(i)}, x^{(j)}) \geq \alpha$, we can approximate it by checking

$$Ham(sig_i \oplus sig_j) \leq \frac{l * d * \arccos \alpha}{\pi}. \quad (13)$$

where $\oplus$ is the XOR bitwise operation and $Ham(\cdot)$ is the Hamming weight of the binary number, which is the number of 1s in the binary number. Since the approximation checking is a bitwise operation, it is usually much faster than the exact calculation of cosine similarity.

### 3.5.2 Parallelization

Our algorithm is inherently parallelizable. First of all, the inference of local expected behavior and global expected behavior are independent and thus they can be computed separately in a trivially parallelized fashion. Moreover, the individual computation of local expected behavior and global expected behavior are also parallelizable. For the former, the examination of contextual neighbor candidates as well as the calculation of the average value of behavioral attributes among the contextual neighbors are independent for each object and can be naturally parallelized. For the latter, the regression model for each behavioral attribute is independent, and therefore, the learning process of each regression model can be parallelized.

## 4. EXPERIMENTS AND ANALYSIS

In this section, we run ROCOD on several datasets and compare the performance with a series of baselines.

### 4.1 Experimental Setup

#### 4.1.1 Dataset and Data Preprocessing

Evaluating outlier detection algorithms is challenging due to lack of publicly available real-world data with ground truth. This problem becomes even more exacerbated when evaluating contextual outlier detection since it requires the data contains contextual and behavioral attributes. In this paper, we employ six datasets to evaluate the performance of our algorithm, two of which contain labeled ground-truth outliers. For the four datasets without ground-truth outliers, we inject contextual outliers using the perturbation scheme described by Song et al. [33] – a de-facto standard for evaluating contextual outlier detection techniques. This scheme works as follows. To inject one outlier into a dataset with $N$ objects, we uniformly select an object $z^{(i)} = (x^{(i)}, y^{(i)})^T$ at random. We then randomly select $p = \min(50, \frac{N}{4})$ objects from the dataset, among which we pick the object $z^{(j)} = (x^{(j)}, y^{(j)})^T$ such that the Euclidean distance between $y^{(i)}$ and $y^{(j)}$ is maximized. We add a new object $z' = (x^{(i)}, y^{(j)})^T$ into the dataset as a contextual outlier [1].

---
[1] In total, we inject $1\% * N$ outliers into the dataset (except for Bodyfat due to the small size of the dataset).

The basic information of the six datasets is shown in Table 1 and is further detailed below.

- **Synthetic**: This dataset is generated using the CAD model [33]. We set the number of contextual attributes and behavioral attributes to be 20, and the number of components for contextual and behavioral attributes to be 16. We randomly generate the centroid and covariance matrix of each component. In total, we generate 50,000 data points and we inject 0.1% contextual outliers using perturbation scheme.
- **Bodyfat**: The dataset is from CMU statlib [2]. We consider attributes on body fat percentage as behavioral attributes and other physical features, e.g., age and chest circumference, as contextual attributes. This dataset does not contain labeled outliers, and we inject outliers to the dataset following the perturbation scheme described above.
- **ElNino**: This dataset is from UCI machine learning repository [3]. It contains oceanographic and the surface meteorological readings taken from a series of buoys positioned throughout the equatorial Pacific. We use the temporal attributes and spatial attributes as contextual attributes (6 attributes) while regard attributes on winds, humidity and temperature as behavioral attributes (5 attributes). In total, we have 93,935 observations and we additionally inject 0.1% contextual outliers as above.
- **Houses**: The dataset comes from CMU statlib. It includes information on house price in California and the relevant variables. We use the house price as the behavioral attributes and other attributes as contextual attributes, such as median income, housing median age, median house value, etc. That dataset contains 20,640 entries and we inject 0.1% contextual outliers as above.
- **YouTube-Twitter**: This dataset is adapted from previous work [1]. It contains both users' information from Twitter and video information from YouTube, where users might re-tweet videos from YouTube. Users' information contains the number of followers and friends, the number of tweets, the fraction of tweets containing YouTube video and so on. It also has video information from YouTube including the number of sharing, the number of views, the number of comments, mean polarization, etc. In this dataset, we intend to detect promotional users whose behavior of sharing videos serves for advertisement. Such users are typically different from the normal population of users. The attributes from Twitter can be naturally adopted as contextual attributes while the attributes from You-Tube can be behavioral attributes. This setting is based on a key intuition that the given Twitter attributes are not directly related to the promotional behavior, yet it is related to the behavior of sharing videos [4]. And the attributes of video shared from YouTube are treated as behavioral attributes since they directly reflect promotion behavior. The dataset is labeled in advance based on the filtering mechanism mentioned in [1] and therefore we have the ground truth of outliers.

- **Kddcup99**: This is a dataset of network connections [5]. It incorporates different types of network intrusions and is widely used for outlier detection. Since we intend to detect intrusions among other connects, we only retain *u2r* and *r2l* attacks and treat them as outliers while removing all other attacks, following the procedure in previous work [28]. Moreover, considering the observation in [41] that *service*, *duration*, *src_bytes* and *dst_bytes* are most essential attributes for intrusion behavior, we use them as behavioral attributes and the rest of attributes are treated as contextual attributes. We take the logarithm of *duration*, *src_bytes* and *dst_bytes* since they exhibit a log-normal distribution as suggested elsewhere [41]. Again in this dataset, we have labeled ground truth outliers.

For each dataset, we also transform categorical attributes into numerical values (adopting 1-of-$m$ encoding technique) and conduct min-max normalization for each attribute.

### 4.1.2 Baselines

We compare ROCOD to the state-of-the-art approaches on outlier detection, including those for detecting contextual outliers. These contain:

- **CAD**: Conditional Anomaly Detection (CAD) proposed by Song *et al.* [33]. It models the contextual attributes and behavioral attributes using a Gaussian Mixture Model but with a probabilistic mapping function to capture how behavioral attributes are related to contextual attributes. In the experiments, we set the number of Gaussian components as 30 (a few alternative values are set but no obvious difference of results is observed). EM algorithm with at most 100 iterations is used to estimate the parameters.
- **LSOD**: Locality Sensitive Outlier Detection [39] is a state-of-the-art representative of a distance based anomaly detection algorithm which calculates an outlierness score of a point based on its distance to its $k$-th nearest neighbor [30]. We set $k$ as 30 after much tuning.
- **LOF**: Local Outlier Factor [6] is a local method that works by comparing the local reachability density of each node to its neighbors. The local density, which can be viewed as a point's context, is defined by referring to the $k$ nearest neighbors and the reachability density produces a more stable measurement. Points with significantly lower density compared to their neighbors are regarded as outliers. In our experiments, we adopt a publicly available implementation [29] and set the range of size of neighborhoods from 10 to 100 as suggested by the authors.
- **COF**: Connectivity-based Outlier Factor [35] differentiates low density from isolation. This method uses a set-based neighborhood paths and is able to detect outliers deviating from a connected pattern. Again, we use a publicly available implementation from the authors [29].
- **GMM**: Gaussian Mixture Model (GMM) [31] models the distributions of the dataset and can be used to measure the outlierness score of objects. Expectation-Maximization (EM) algorithm is usually adopted to estimate the parameters and the outlier scores are obtained by looking at the probability density under the distribution. Here, we set the number of Gaussian components as 30 and the maximum number of iterations in EM algorithm as 100.

For ROCOD, the cosine similarity threshold $\alpha$ is chosen by looking at the distribution of randomly sampled pairs of objects. In total, there are $\frac{N(N-1)}{2}$ pairs of objects in a dataset with size $N$. We uniformly sample $r$ pairs of them and calculate their cosine similarities. When ranked in decreasing order, we pick the 5-th percentile as the cosine similarity threshold. For the local expected behavior, we leverage LSH to find out the contextual neighbors. We adopt both linear and non-linear models for the global expected behavior. Specifically, we choose Ridge regression for the linear model and tree regression [5] for the nonlinear model, denoted as `ROCOD_1` and `ROCOD_2` respectively. As comparisons, we also use only local expected behavior and global expected behavior to flag outliers, denoted as **LEB** and **GEB**.

---
[2] ftp://rcom.univie.ac.at/mirrors/lib.stat.cmu.edu/
[3] https://archive.ics.uci.edu/ml/datasets/El+Nino
[4] Many promotional users share videos actively even though they have few original tweets and comments. But if a user posts a huge amount of tweets and obtain many comments, then his/her behavior of sharing many videos might not be a sign of a being promotional user.
[5] http://kdd.ics.uci.edu/databases/kddcup99/task.html

| Datasets | Outliers | # Objects $N$ | # Outliers | Contextual Attributes | Behavior Attributes |
|---|---|---|---|---|---|
| Synthetic | Injected by perturbation scheme. | 50,500 | 500 | 20 | 20 |
| Bodyfat | Injected by perturbation scheme. | 277 | 25 | 13 | 2 |
| ElNino | Injected by perturbation scheme. | 94,874 | 939 | 6 | 5 |
| Houses | Injected by perturbation scheme. | 20,846 | 206 | 8 | 1 |
| YouTube-Twitter | Promotional users. | 62,458 | 2,974 | 48 | 41 |
| Kddcup99 | u2r and r2l attacks. | 98,372 | 1,094 | 47 | 69 |

Table 1: Basic information of the datasets. Outliers are injected by perturbation scheme for datasets without ground truth.

| | | | | | | | | | |
|---|---|---|---|---|---|---|---|---|---|
| | | | | Synthetic | | | | | |
| **Metrics** | `ROCOD_1` | `ROCOD_2` | **LEB** | **GEB** | **CAD** | **GMM** | **LSOD** | **LOF** | **COF** |
| PRC (AUC) | 0.392 | **0.913** | 0.781 | 0.679 | 0.628 | 0.055 | 0.265 | 0.384 | 0.135 |
| Top-100 Precision | 0.740 | **0.950** | 0.890 | 0.860 | 0.880 | 0.150 | 0.450 | 0.710 | 0.230 |
| Top-100 nDCG | 0.749 | **0.960** | 0.867 | 0.877 | 0.891 | 0.192 | 0.431 | 0.730 | 0.280 |
| | | | | Bodyfat | | | | | |
| **Metrics** | `ROCOD_1` | `ROCOD_2` | **LEB** | **GEB** | **CAD** | **GMM** | **LSOD** | **LOF** | **COF** |
| PRC (AUC) | **0.750** | **0.750** | 0.614 | **0.750** | **0.750** | 0.430 | 0.644 | 0.725 | 0.667 |
| Top-10 Precision | 0.900 | **1.000** | 0.800 | 0.900 | 0.900 | 0.400 | 0.300 | 0.400 | 0.300 |
| Top-10 nDCG | 0.936 | **1.000** | 0.875 | 0.933 | 0.936 | 0.287 | 0.206 | 0.305 | 0.215 |
| | | | | ElNino | | | | | |
| **Metrics** | `ROCOD_1` | `ROCOD_2` | **LEB** | **GEB** | **CAD** | **GMM** | **LSOD** | **LOF** | **COF** |
| PRC (AUC) | 0.670 | **0.990** | 0.964 | 0.883 | 0.220 | 0.461 | 0.806 | 0.797 | 0.767 |
| Top-100 Precision | 0.960 | **1.000** | 0.990 | **1.000** | 0.400 | 0.790 | **1.000** | 0.950 | **1.000** |
| Top-100 nDCG | 0.970 | **1.000** | 0.968 | **1.000** | 0.404 | 0.825 | **1.000** | 0.947 | **1.000** |
| | | | | Houses | | | | | |
| **Metrics** | `ROCOD_1` | `ROCOD_2` | **LEB** | **GEB** | **CAD** | **GMM** | **LSOD** | **LOF** | **COF** |
| PRC (AUC) | 0.656 | **0.766** | 0.312 | 0.634 | 0.232 | 0.135 | 0.101 | 0.116 | 0.119 |
| Top-100 Precision | 0.740 | **0.840** | 0.300 | 0.650 | 0.350 | 0.210 | 0.080 | 0.260 | 0.250 |
| Top-100 nDCG | 0.694 | **0.860** | 0.226 | 0.717 | 0.461 | 0.222 | 0.067 | 0.247 | 0.245 |
| | | | | YouTube-Twitter | | | | | |
| **Metrics** | `ROCOD_1` | `ROCOD_2` | **LEB** | **GEB** | **CAD** | **GMM** | **LSOD** | **LOF** | **COF** |
| PRC (AUC) | 0.141 | **0.146** | 0.136 | 0.124 | 0.125 | 0.131 | 0.124 | 0.105 | 0.106 |
| Top-100 Precision | **0.530** | 0.470 | 0.440 | 0.360 | 0.370 | 0.180 | 0.280 | 0.230 | 0.260 |
| Top-100 nDCG | **0.605** | 0.440 | 0.426 | 0.413 | 0.362 | 0.168 | 0.272 | 0.223 | 0.240 |
| | | | | KDDcup99 | | | | | |
| **Metrics** | `ROCOD_1` | `ROCOD_2` | **LEB** | **GEB** | **CAD** | **GMM** | **LSOD** | **LOF** | **COF** |
| PRC (AUC) | 0.137 | **0.143** | 0.071 | 0.051 | 0.027 | 0.128 | 0.027 | 0.019 | 0.014 |
| Top-100 Precision | 0.390 | **0.600** | 0.070 | 0.020 | 0.300 | 0.000 | 0.070 | 0.020 | 0.000 |
| Top-100 nDCG | 0.293 | **0.518** | 0.129 | 0.031 | 0.316 | 0.000 | 0.085 | 0.015 | 0.000 |

Table 2: Performance comparisons of baselines on 6 datasets. `ROCOD_1` uses the linear model in the global expected behavior while `ROCOD_2` adopts the non-linear model. LEB and GEB are the approaches utilizing only local expected behavior and global expected behavior (with non-linear model) respectively. Three metrics are used to evaluate the performance (higher is better). Best performances w.r.t. each metric are shown in bold.

| Datasets | `ROCOD_1` | `ROCOD_2` | **LEB** | **GEB** | **CAD** | **GMM** | **LSOD** | **LOF** | **COF** |
|---|---|---|---|---|---|---|---|---|---|
| Synthetic | 17.7 | 140.5 | 15.1 | 124.3 | 6,710.8 | 1,712.5 | 326.2 | 263.0 | 178.0 |
| ElNino | 35.6 | 40.4 | 35.2 | 5.0 | 1,378.4 | 472.6 | 177.2 | 929.0 | 317.0 |
| Houses | 3.0 | 3.2 | 2.8 | 0.3 | 2,447.3 | 87.0 | 7.7 | 109.0 | 29.0 |
| YouTube-Twitter | 55.8 | 179.8 | 50.1 | 140.1 | 16,486.2 | 9,283.7 | 977.9 | 463.0 | 350.0 |
| KDDcup99 | 613.0 | 624.9 | 594.5 | 33.3 | 24,036.3 | 9,126.0 | 1,166.3 | 1,259.0 | 809.0 |

Table 3: Running time comparisons (in seconds). All the methods are executed using one process.

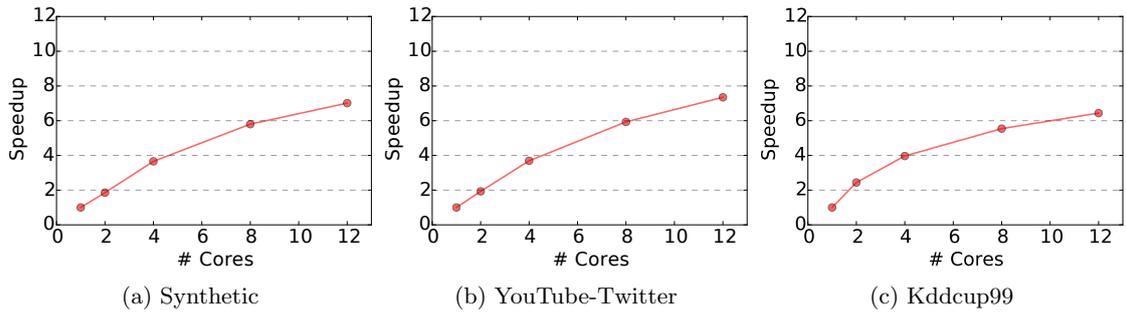

Figure 2: Speedups of `ROCOD` on three larger datasets.

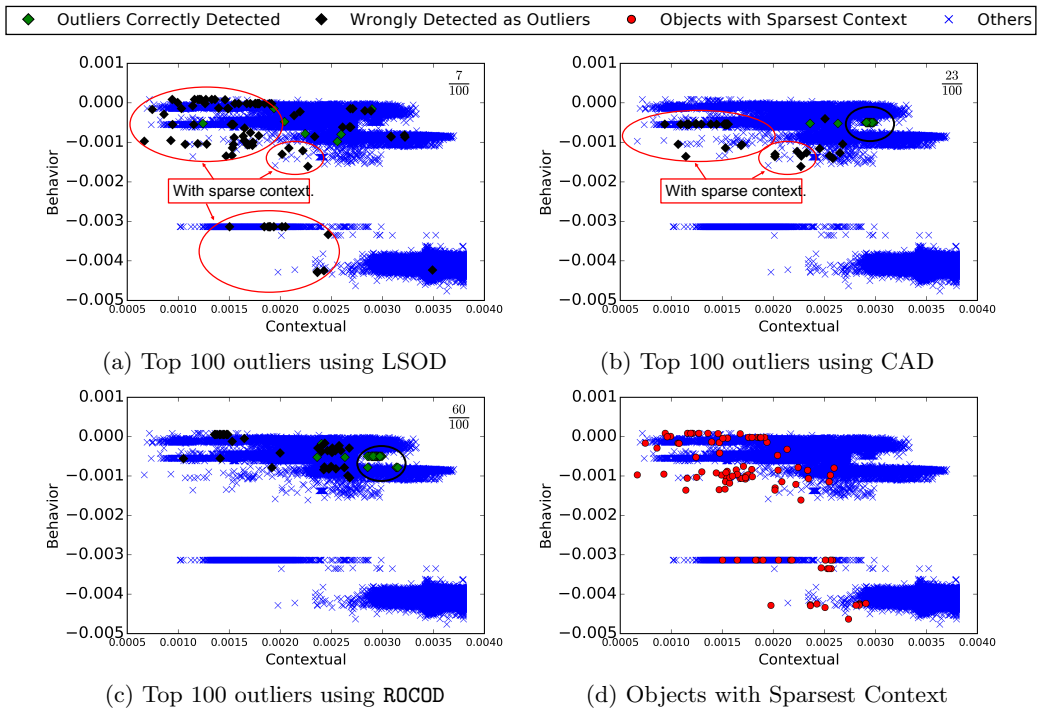

Figure 3: Visualization of data points of Kddcup99 in the 2-D coordinate. (a)-(c) show outliers detected by different approaches. Red round circles highlight objects with sparsest contexts in (a) and (b). Black round circles highlight clusters of outliers correctly detected by CAD and `ROCOD`. (d) shows 100 objects with the sparsest contextual attributes, indicated by red dots.

### 4.1.3 Evaluation metrics

All the outlier detection approaches considered in this paper output a full list of objects ranked by their outlierness scores with higher ones on the top. To report $n$ outliers, these approaches will return the $n$ objects on the top of the ranked list. In some real-world applications, e.g., credit card fraud detection, one might focus more on the overall performance of precision and recall. In other applications, however, one might try to avoid information overload and focus only on the top of the list. Considering the requirements of various applications we consider three broad metrics to comprehensively evaluate the approaches.

**Precision-Recall Curve**: Precision-Recall curve (PRC) is obtained by plotting precision ($y$-axis) versus recall ($x$-axis) for all possible choices of $n$. We use PRC instead of Receiver Operating Characteristic (ROC) since ROC can present an overly optimistic result of an algorithm and is less informative in our problem, where the class distribution is highly skewed [12]. In this experiment, we summarize PRC using area under the curve.

**Top-$n$ Precision**: It is defined as the precision of the objects ranked among top-$n$, which is also called *precision at $n$*. To calculate it, we examine the ratio of correctly detected outliers among $n$ objects ranked on the top. Formally,

$$Precision(n) = \frac{|\{i | i \in O \& rank(i) \leq n\}|}{n}, \quad (14)$$

where $O$ is the set of outliers from ground-truth.

**Top-$n$ nDCG**: Discounted cumulative gain (DCG) is commonly used in information retrieval to measure the effectiveness of ranking [19]. Different from top-$n$ precision, it uses a graded relevance scale to evaluate the gain based on the position of each object. We adopt the normalized DCG [19] as follows.

$$nDCG(n) = \frac{DCG(n)}{IDCG(n)}, \quad DCG(n) = \mathbb{1}_O(1) + \sum_{i=2}^{n} \frac{\mathbb{1}_O(i)}{\log_2 i}. \quad (15)$$

Here, $\mathbb{1}_O(i)$ is an indicator function, i.e. 1 if $i$ is an outlier and 0 otherwise. $IDCG(n)$ is the ideal DCG when all the top-$n$ objects are outliers (indicator function values will all be 1).

Moreover, we also measure the efficiency of each approach. We report the **wall clock running time** of each algorithm.

## 4.2 Experiment Results and Analysis

We run the experiments on a Linux Machine with two Intel Xeon x5650 2.67GHz CPUs. It contains 12 cores and 48GB of RAM. All the algorithms are implemented in `C++` and compiled using Intel compiler. `OpenMP` is used to exploit the parallelism. Results are shown in Table 2, Table 3, and Figure 2. Table 2 shows the performance of each approach on detecting outliers in all the datasets, measured by three different metrics. We highlight some observations below.

1) We can see that `ROCOD` (either `ROCOD_1` or `ROCOD_2`) performs the best at almost all the datasets. In some datasets, especially Synthetic, Houses and KDDcup99, `ROCOD` significantly outperforms other baselines on all three evaluation metrics. For example, the top-100 precision of `ROCOD_2` on KDDcup99 is twice as much as the best of baselines (CAD). The advantage of `ROCOD` is more pronounced in terms of top-$n$ metrics, indicating our method is able to show outliers at the top more precisely. In Synthetic and ElNino dataset, the top-100 precision and nDCG of our approach is almost perfect (very close to 1.0).

2) Without separating contextual and behavioral attributes, general outlier detection approaches (e.g., LSOD, LOF and COF), perform poorly on the datasets. This issue is more evident on Bodyfat and Houses dataset, which contain more contextual attributes than behavioral attributes. The main reason is that these approaches simply combine contextual attributes with behavioral attributes and the effect of contextual attributes on outlierness score may obfuscate the role of behavioral attributes. If one observes the Houses dataset as an example, only 1 out of 9 attributes is behavioral attribute, simply combining the attributes leads to very poor performance (see the performance of LSOD and LOF in this dataset). Similar phenomenon can be observed in Kddcup99 dataset, where contextual outlier detection techniques CAD and `ROCOD` achieve top-100 precision as large as 60% while others mostly cannot obtain higher than 10%.

3) `ROCOD` with the non-linear model for global expected behavior (`ROCOD_2`) outperforms the one with the linear model (`ROCOD_1`). This is not surprising and consistent with our intuition that the non-linear model is more capable of modeling the complex relationship among attributes. In fact, `ROCOD_2` obtains the best performance in all the datasets except YouTube-Twitter, where `ROCOD_1` performs the best on top-100 precision and nDCG while `ROCOD_2` is better in terms of AUC of the precision-recall curve.

We also compare the running time of all the baselines. Table 3 presents the detailed running time of each method with one process. From the table, we can see that `ROCOD` runs much faster than other baselines on all the datasets. Even with the non-linear model, `ROCOD_2` is still 40X faster than the state-of-the-art contextual outlier detection method CAD, and is slightly faster than LSOD and LOF. We also observe that `ROCOD_1` performs faster than `ROCOD_2`.

**Benefits of Scalability Optimization:** Additionally, to drill down on the benefits of the LSH optimization for estimating the contextual neighbors, we run `ROCOD_2` without using LSH on both YouTube-Twitter and KDDcup99 datasets (results on other datasets show similar benefits). `ROCOD_2` takes 674.2 and 1083.8 seconds respectively on the two datasets – significantly longer than 179.8 and 624.9 seconds when using LSH as shown in Table 3 – demonstrating that LSH does help improve the efficiency of `ROCOD` drastically. We also illustrate the benefits of parallelizing `ROCOD` as described in Section 3.5 and run it on the three larger datasets (non-linear model is used for global expected behavior). Figure 2 shows the speedup of `ROCOD` on these three datasets (observing a speedup of up to 8 on 12 threads on YouTube-Twitter).

## 4.3 Drilling Down on Efficacy Gains

Here, we drill-down to distil the performance gains of `ROCOD`. We take Kddcup99 dataset as an example and visualize outliers flagged by different methods among all other objects in 2-D coordinates. In order to visualize high-dimensional data, we adopt the typical method of extracting the principal component. For the Kddcup99 dataset, we extract the largest component from the contextual attributes space and behavioral attributes space respectively and plot the data points directly in the 2-D coordinate. Figure 3 shows the visualization of all the objects. Green diamonds are outliers correctly identified while black diamonds are normal objects but flagged incorrectly as outliers by the approaches. We show the top-100 outliers detected by each method and the precision is shown at the upper right corner of each plot. We include the results from LSOD, CAD and `ROCOD`, which are the three best approaches in this dataset. Moreover, we also identify 100 objects with sparsest contexts, i.e., their contextual attributes are very different from others. We select top 100 objects and mark them using red dots in Figure 3d.

Comparing Figure 3c with Figure 3a and Figure 3b, we notice that LSOD and CAD tend to mistakenly detect similar groups of normal objects as outliers, which are highlighted in red circles in the plots (Figure 3a and Figure 3b), while `ROCOD` avoids similar mistakes. To understand the reason for this observation, we look at these groups of objects in Figure 3d and find out that most of them are show in red dots and therefore are objects containing sparse contextual attributes. This strongly supports our statement that existing outlier detection techniques (such as LSOD and CAD) tend to assign higher outlierness scores to objects with anomalous contextual attributes though they are normal considering their behavioral attributes. Even state-of-the-art approach CAD cannot properly resolve this issue. `ROCOD`, however, is not affected much by these objects and is able to impartially measure the outlierness scores of them. Moreover, it is also interesting to observe that though ROCOD and CAD are two totally different approaches, they correctly detect the same group of outliers, circled by black bold rectangles in Figure 3b and 3c. In general, ROCOD is much better than CAD at correctly identifying outliers in this dataset.

## 5. CONCLUSION AND FUTURE WORK

We propose a comprehensive approach ROCOD to exploit contextual attributes for detecting outliers, particularly dealing with the existing issue caused by the sparsity of context. We introduce local expected behavior and global expected behavior models to infer the behavioral attributes and describe a natural algorithm, ROCOD, to fuse them. We also develop some critical optimizations for scaling up the algorithm. Experimental results show that ROCOD detects outliers more accurately and efficiently than prior approaches.

As part of future work, we plan to investigate how to automatically extract behavioral attributes and contextual attributes from such datasets. This is a non-trivial problem often requiring some domain insight. Of particular interest is to detect such outliers in data collected from a gaming platform designed for physical rehabilitation of stroke patients [24].

**Acknowledgements.** This work is supported by NSF Award NSF-EAR-1520870 and NSF-DMS-1418265.